\documentclass[preprints,article,accept,pdftex,moreauthors]{Definitions/mdpi} 

\firstpage{1} 
\pubvolume{1}
\issuenum{1}
\articlenumber{0}
\pubyear{2024}
\copyrightyear{}
\datereceived{ } 
\daterevised{ } 
\dateaccepted{ } 
\datepublished{ } 
\hreflink{https://doi.org/} 
\pdfoutput=1 

\Title{Optimizing Josephson Junction Reproducibility in 30 kV E-beam Lithography: Analysis of Backscattered Electron Distribution}
\TitleCitation{Title}

\Author{A. M. Rebello$^{1\ddagger}$\orcidA{0000-0003-2797-3470}, L. M. Ruela$^{2\ddagger}$\orcidA{0000-0001-5052-5778}, G. Moreto$^{2}$\orcidA{0009-0009-8493-8349}, N. Y. Klein$^{1}$\orcidA{0000-0002-1209-568X}, E. Martins$^{1}$\orcidA{0000-0003-4562-7205}, I.  S.  Oliveira$^{1}$\orcidA{0000-0002-6970-2057}, J. P. Sinnecker$^{1}$\orcidA{0000-0001-5211-901X}, F. Rouxinol$^{2\ddagger}$\orcidA{0000-0003-2899-5364}}

\AuthorNames{A. M. Rebello, L. M. Ruela, G. Moreto, N. Y. Klein, E. Martins, I.  S.  Oliveira, F. Rouxinol, J. P. Sinnecker}

\AuthorCitation{Lastname, F.; Lastname, F.; Lastname, F.}

\address{%
$^{1}$ \quad Coordenação de Matéria Condensada, Física Aplicada e Nanociência(COMAN). Centro Brasileiro de Pesquisas Físicas(CBPF), Rio de Janeiro, RJ, Brazil\\
$^{2}$ \quad Universidade Estadual de Campinas (Unicamp), IFGW, Quantum Device Physics Laboratory, SP, Brazil\\
$^{3}$ \quad Leopoldo Américo Miguez de Mello Research, Development and Innovation Center (CENPES), RJ, Brazil}

\corres{Correspondence: artrebello@cbpf.br,rouxinol@unicamp.br}

\firstnote{These authors contributed equally to this work.}

\abstract{This paper explores methods to enhance the reproducibility of Josephson junctions, crucial elements in superconducting quantum technologies, when employing the Dolan technique in 30 kV e-beam processes. The study explores the influence of dose distribution along the bridge area on reproducibility, addressing challenges related to fabrication sensitivity. Experimental methods include E-beam lithography, with electron trajectory simulations shedding light on backscattered electron behavior. We demonstrate the fabrication of different junction geometries, revealing that some geometries significantly improve reproducibility by resulting in a more homogeneous dose distribution over the junction area.}


\begin{document}


\section{Introduction}

Superconducting Quantum computing represents a frontier in contemporary physics and engineering, promising revolutionary advancements in computation\cite{google,20q},  communication\cite{com,crypt} and sensing\cite{walsh_2021,sense}. Central to these technologies are Josephson junctions, critical components that enable the unique nonlinear properties of superconducting quantum circuits. Much progress has been made to improve functionality and quality of Josephson junctions and their applications \cite{dieletric,lowdissipation}. The fabrication process, however, remains exceedingly sensitive; even minor variations in junction area or stochastic fluctuations in oxide formation can lead to inconsistencies in the overall oxide barrier, making the reproducibility of Josephson junctions a hindering factor in the performance and scalability of quantum devices \cite{materials}. This letter aims to enhance the stability of the fabrication process by addressing the impact of backscattered electrons in the design of Josephson junctions, for low-energy electron beam lithography (EBL).

\begin{figure*}[]
\resizebox{\hsize}{!}{\includegraphics[clip=true]{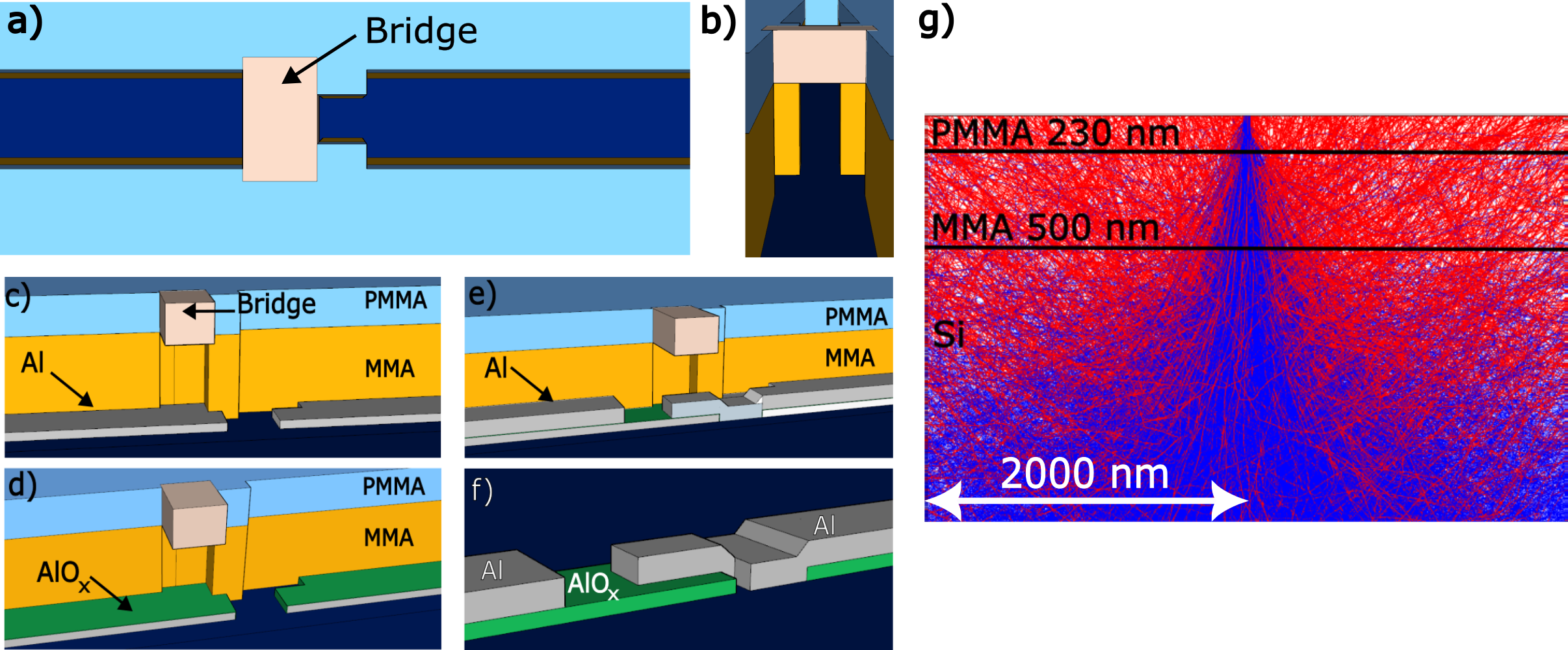}}
\caption{Schematic representation of the pattern transferred onto a double-layer resist stack followed by aluminium deposition process, delineating the steps involved in fabricating Josephson junctions,  alongside electron trajectory simulation within the resist layers. a) Dolan Josephson junction scheme showcasing the exposed area (in dark blue) and the central bridge region (in grey). b) Bird's-eye perspective of the anticipated Josephson junction bridge structure. c) Initial $30^\circ$ angle deposition, d) oxidation phase, e) subsequent -$30^\circ$ angle deposition. f) Representation of the Josephson junction post-lift-off process, with the green coating symbolizing the $AlO_x$ layer. g) Visualization of electron dispersion trajectories in a 230 nm PMMA layer (top) and a 500 nm MMA Co-polymer layer (bottom), both situated on a silicon substrate, under the influence of a 30 kV electron beam. The trajectories of the primary electrons from the incident beam are depicted in blue, whereas the backscattered electrons are illustrated in red.}\label{img1}
\end{figure*}

In this paper, we present modifications to the fabrication methodology of Josephson junctions using 30 kV e-beam lithography, specifically employing the Dolan technique\cite{Dolan,schmidlin, osman, sterr}. While the technical aspects of the 30 kV e-beam process are well-documented, bridge structure integrity is a known issue in fabricating these structures. Some groups claim robustness is compromised due to stress which occurs to PMMA layer depending on geometry used \cite{stress,kellystress}. Others deal with this by pre-exposing the bottom resist \cite{doubleexpose}. Our study addresses issue by focusing on enhancing fabrication reproducibility through the strategic selection of optimal geometries which correctly engineer the doses of backscattered electrons on the bridge area. Moreover, additional teams have recognized the issue we raise in this letter, they have attempted to apply complex commercially available 3D proximity effect correction (PEC)\cite{3dpec}. However even at 100kV where the backscattered electron distribution is nearly homogeneous they had to resort to manually modifying the dose in different parts of their exposure layout to achieve desired results. We do acknowledge that even the traditional PEC \cite{pec} improved our results, and PEC was in fact used for the fabrication of all Junctions in our study, nevertheless it was not enough to yield satisfying results. We will show how backscattered electron distributions can be engineered to effectively yield more robust fabrication processes. Such advancements are crucial for ensuring the reliability of Josephson junction within superconducting quantum circuits, particularly when utilizing low-energy 30 kV EBL.

\begin{figure*}[]
\begin{adjustwidth}{-\extralength}{0cm}
\centering
\includegraphics[width=15.5cm]{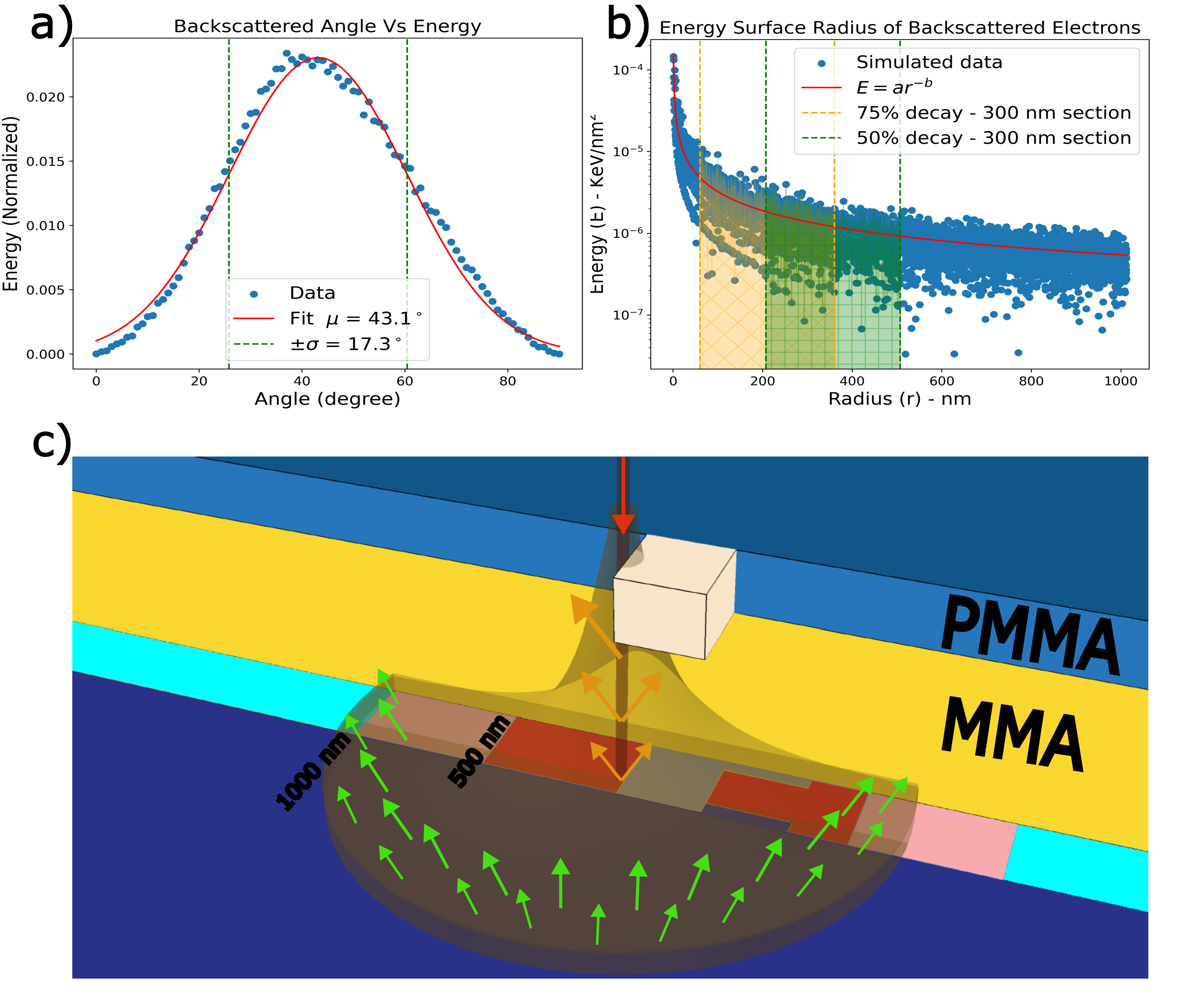}
\end{adjustwidth}
\caption{Statistical analysis of simulated backscattered electron trajectories illustrating the correlation with angle and radius, accompanied by a schematic representation of the energy surface. a) Simulated backscattered energy versus angle, with the green dashed line indicating the 2$\sigma$ confidence region. b) Radius of the deposited energy surface for backscattered electrons as determined by simulation. First 300 nm section to be within resist material selectivity threshold is from 60 to 360 nm where the fitted backscattered energy will decay 75$\%$ from start to end, region is shaded in orange. Comparative 50$\%$ decay region in shaded in green. c) The energy surface of backscattered electrons (orange shade) surrounding the incident beam (red), integrated into a cross-sectional diagram of the bridge region. The pattern areas within 500 nm of the bridge section are highlighted in red, and areas within 1000 nm are shown in light red. Red arrow depicts the beam incident direction, orange arrow shows electrons backscattered within the resist stack, green arrows indicate backscattered electrons permeating from within the substrate.}
\label{img2}
\end{figure*}

\section{Materials and Methods}

Fundamentally, the objective is to pattern the resist to form a bridge, as illustrated in Figure \ref{img1}a-d. As depicted, the challenge involves employing a straight electron beam to intricately sculpt sideways beneath the resist's top layer. This process is feasible because the electron beam generates backscattered electrons along its path and it penetrates several micrometers into the substrate, implying backscattered can travel a long distance from point of incidence. Backscattered electrons are highlighted in Figure \ref{img1}g with red trajectories. A critical aspect of enhancing the fabrication process quality is understanding the impact of these backscattered electrons and acknowledging the variability in the sensitivity of the resist layers employed, specifically PMMA (230 nm) over MMA (500 nm). 

The challenge in electron beam lithography (EBL) processes traditionally lies in optimizing to achieve the best matching feature dimensions. For any given exposure area, a portion of the incident dose is inevitably distributed to surrounding regions, resulting in variations in the deposited dose near the beam. This phenomenon, known as the proximity effect, is mitigated using Proximity Effect Correction (PEC) techniques, which modify the dose distribution across the features to achieve a constant deposition dose within the exposed area \cite{pec}. However, these techniques primarily aim to establish a uniform dose for a given design. Here, we endeavour to advance further by examining how undercuts can be engineered to preserve the design structures of the upper resist layer. The challenge emerges from the discrepancy between the dose required to pattern the design and the dose from backscattered electrons needed to sensitize the lower resist layer, all while maintaining the integrity of the top resist layer. Given that patterning the top resist layer usually does not deposit enough dose to open the undercut in the bottom layer, an increased base dose is necessary. This adjustment, however, leads to overdosing in both the top and bottom resist layers, potentially causing deformation, weakening, and deviation from designed dimensions.   While PEC helps prevent damage to the bridge structure, it is often insufficient for 30 kV applications in fully address these issues. Thus, as demonstrated in our study, a careful selection of geometric features is also crucial. We show that this approach is of uttermost importance in creating reproducible structures for the fabrication of Josephson junctions.

\begin{figure*}[]
\begin{adjustwidth}{-\extralength}{0cm}
\centering
\includegraphics[width=16cm]{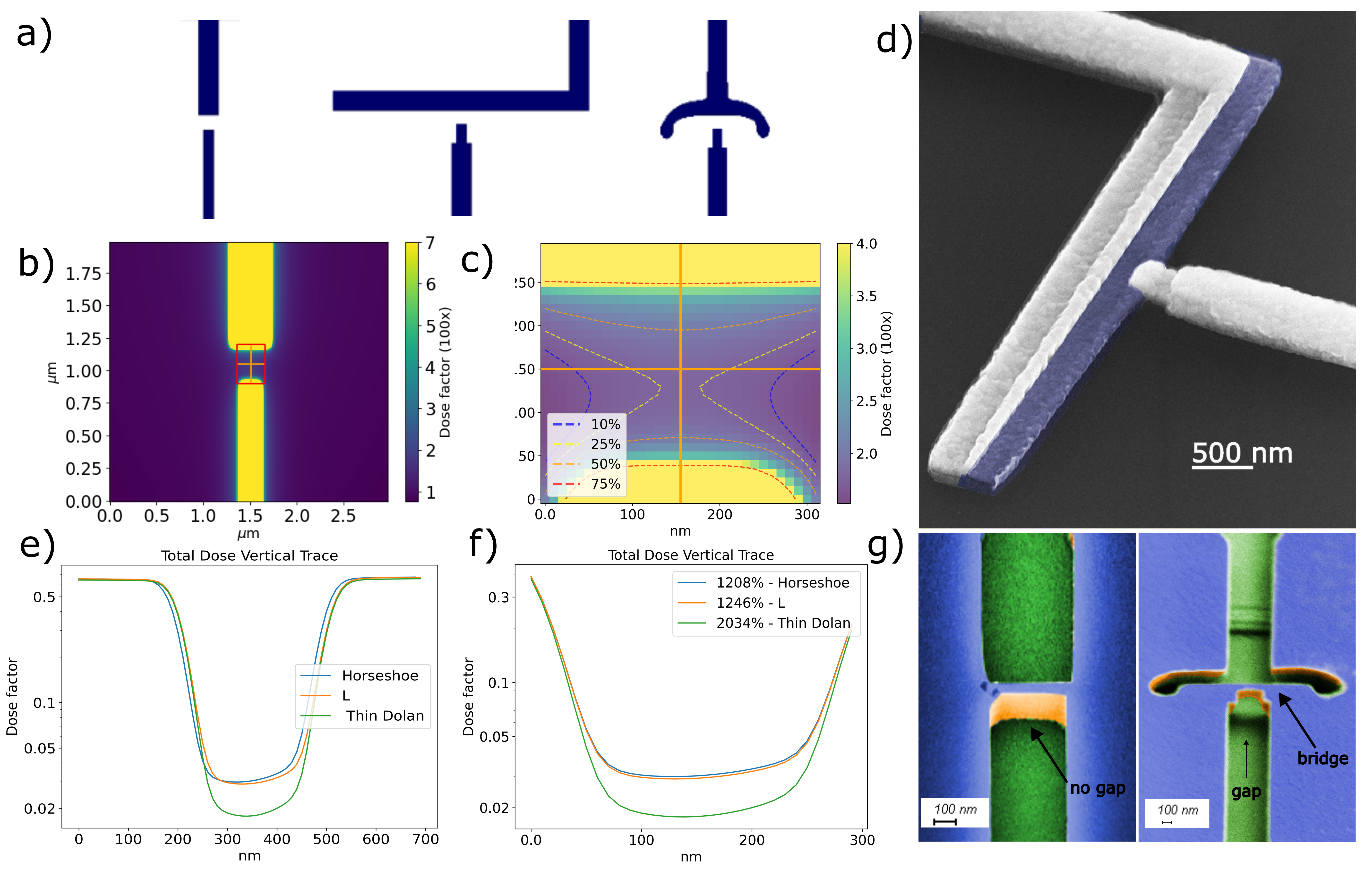}
\end{adjustwidth}

\caption{a) Designs employed to investigate the effects of backscattered electrons across different geometries, identified from left to right as: thin Dolan, L, and horseshoe junction designs; b) Resulting dose map from the integration of the Point Spread Function (PSF) over the thin geometry, with detailed analyses presented in panels (e) and (f). c) Detailed view of the unexposed bridge region, with percentiles marking the total deposited dose per region; (d) Colored scanning electron microscope (SEM) image of an L-shaped Josephson junction, where the blue region indicates the first deposited aluminium layer, and the $AlO_x$ tunnelling barrier is highlighted in the center; e) Total dose distribution profiles along vertical trace, including some of exposed region (200 nm on each side), f) only 300 nm unexposed section, for horseshoe, L and Thin Dolan geometry - percentiles here denote the range of maximum dose variation. (g) Angled coloured SEM images showcasing the resist stack; on the left, the Thin Dolan pattern is inscribed without bridge formation, whereas on the right, the horseshoe pattern is exposed, clearly displaying the bridge structure. Green indicates Si substrate, blue PMMA resist surface and orange for resist side walls seen at an angle. }
\label{img3}
\end{figure*}

The Josephson junctions were fabricated through a single e-beam lithography process followed by a standard Dolan-bridge double-angle evaporation of aluminium in a dedicated ultra-high vacuum (UHV) deposition system\cite{Dolan}. This system, crucial for finely adjusting the tunnelling barrier and preventing contamination, operates under conditions that significantly enhance the quality and reproducibility of the junctions. A detailed description of sample fabrication and techniques used is provided in the appendix. Briefly, the fabrication process begins with e-beam exposure, followed by development in MIBK:IPA 1:3 and rinsing in IPA to stop the development process, then gently drying with N$_2$. The samples are subsequently placed in the deposition chamber for thin film deposition and oxidation. The first step involves depositing ultrapure (99.999\%) Al at a 30$^\circ$ angle relative to the sample's normal (Figure \ref{img1}c), using e-beam vapour deposition in a UHV environment ($10^{-9}$ Torr) to minimize impurities and achieve anisotropic deposition by positioning the sample approximately a meter away from the crucible. Following the first deposition, the sample is moved to a separate chamber and exposed to an $O_2$ atmosphere at 5.7 Torr to form a controlled oxide barrier (Figure \ref{img1}d). Finally, the sample is returned to the main chamber for a second Al deposition at an angle of -30$^\circ$ (Figure \ref{img1}e), completing the fabrication process.

Following the deposition process, the sample is subjected to a lift-off procedure, uncovering the structures depicted in Figure \ref{img1}f. The thin oxide barrier formed during deposition exhibits a notable variation in characteristic resistance at room temperature, which can range from a few Ohms to tens of kOhms, depending on the junction's area and size. This variation in room temperature resistance is directly proportional to the critical current value below the superconductor's critical temperature. By optimizing this resistance, it is possible to finely adjust the performance of these devices, enhancing their functionality and efficiency in superconducting circuits.

Achieving consistent resistance measurements for multiple junctions at room temperature necessitates adherence to a meticulously optimized fabrication protocol. Despite such rigorous optimization, the inherent physical variations on the fabrication process can still lead to fluctuations in resistance measurements \cite{schmidlin}, highlighting the fragile nature of this method and underscoring the imperative need for solutions aimed at enhancing reproducibility, emphasizing the importance of both precision in the fabrication process and the pursuit of innovative strategies to ensure the reliability of Josephson junctions.

Having delineated the fabrication process, we now turn our attention to the optimization of electron beam lithography (EBL) through the simulation of electron trajectories. To elucidate the interaction between the electron beam and our sample, we employed Casino \cite{cas} software for simulating the trajectories of electrons. Utilizing the Monte Carlo method for these simulations allows us to integrate the findings directly into our fabrication strategy. In the software, we define a bi-layer resist over a Si substrate and their respective material properties. The initial layer is composed of an MMA co-polymer, with a density of \(0.80\ g/cm^3\) and a thickness of \(500\ nm\), followed by a second layer of PMMA, having a density of \(1.14\ g/cm^3\) and a thickness of \(230\ nm\). The chosen substrate is silicon (Si), with a density of \(2.33\ g/cm^3\). Data was gathered through the simulation of \(2\) million electron trajectories, employing a beam radius of \(10\ nm\) and beam energy of \(30\ kV\). 

\section{Results}

To elucidate the role of backscattered electrons, the distribution of energy was analyzed with respect to both the scattering angle, as depicted in Figure \ref{img2}a, and the radius of energy distribution, as shown in Figure \ref{img2}b. The analysis revealed that the scattering angle's energy deposition is best modeled by a normal distribution, with the most probable scattering angle centered around \(\mu \approx 43 \pm 17^\circ\). However, electron scattering occurs throughout the beam's path, necessitating an examination of the cumulative effect of this scattering, represented by the radius of the backscattered energy surface. Through fitting the data to a power-law decay, described by the relationship \(\text{Energy} = a \cdot \text{Radius}^{-b}\), we derived the equation E(r) =  \(1.13 \times 10^{-4} \cdot r^{-0.77}\) to characterize the energy distribution's decay. 

This distribution addresses scattering effects within the resist and substrate that spatially and energetically redistribute the electrons of the incident beam. As lower energy particles exhibit a higher scattering probability, more scattering occurs within the resist and substrate along the beam's path. Farther from the beam, distribution becomes approximately constant. We show this effect plotting the backscattered energy surface distribution in three dimensions as a simplified model, aligning it with the cross-section of the exposure pattern diagram. This method qualitatively displays the cumulative energy distribution’s decay along the junction area, as shown in Figure 2c.

\begin{figure*}[]
\begin{adjustwidth}{-\extralength}{0cm}
\centering
\includegraphics[width=15.5cm]{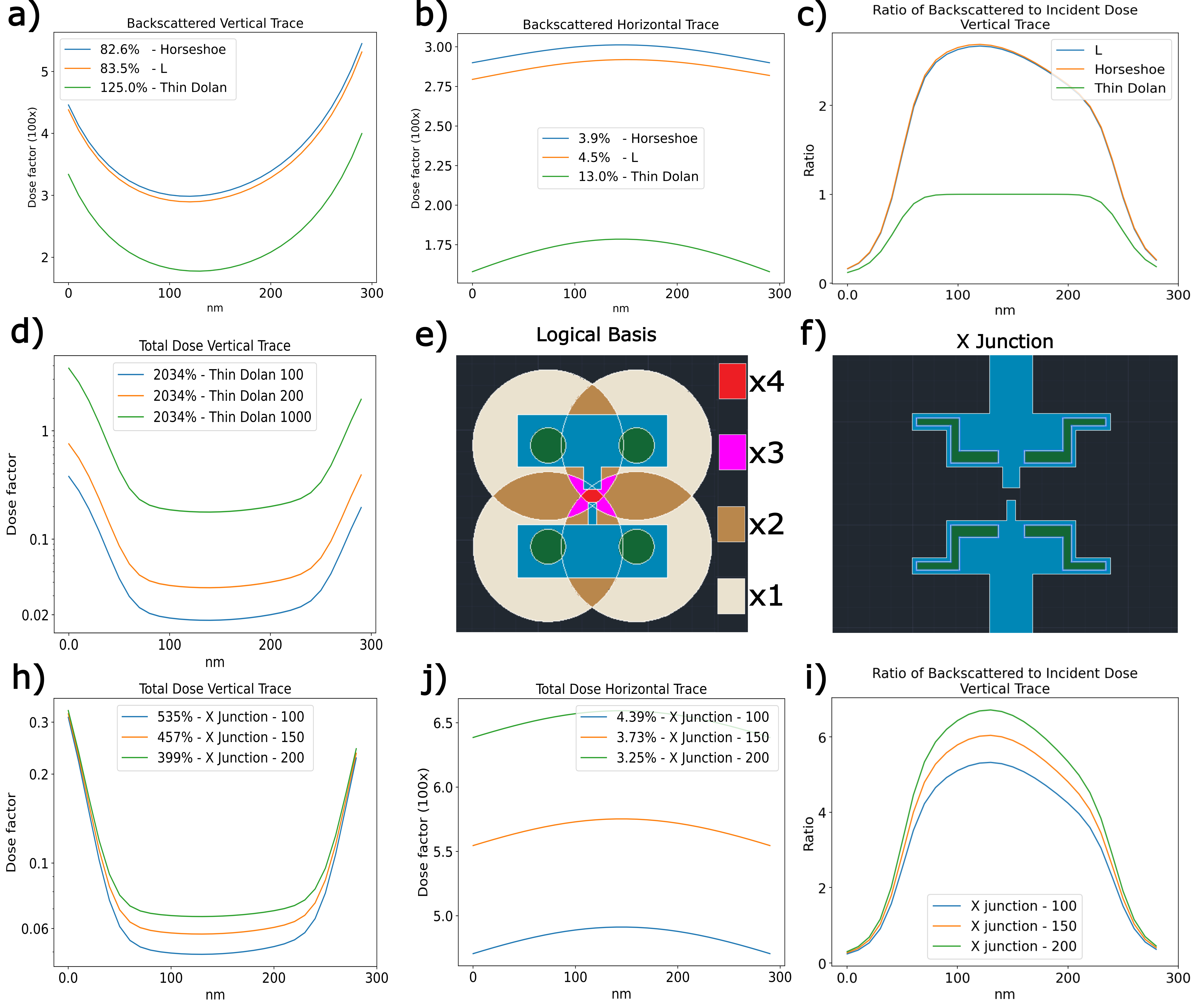}
\end{adjustwidth}
\caption{Analysis of dose factors and the ratio of backscattered to total incident dose for various geometries investigated in this study, alongside a proposed geometry designed to utilize backscattered electrons for undercut definition. The top three figures depict: a) Distribution of backscattered electrons along a vertical trace; b) Distribution along a horizontal trace; c) Ratio of backscattered to incident dose over the vertical trace. d) Dose variation observed in a thin Dolan geometry. e) The logical basis to create a new geometry, specifically conceived to tailor the distribution of backscattered electrons, thereby minimizing variance across the junction area. The blue region is intended to receive the minimal necessary dose to develop the top resist layer, with backscattered electrons being generated within the green circle region by a higher dose factor. The larger circles represent a simplified model for the overlap of backscattered regions, assuming point sources, with different colours indicating the degree of overlap. f) X junction Geometry designed with features to retain geometric resolution while achieving the (4:1) ratio for the h) total deposited dose over the vertical trace.  j) Horizontal profile of the total deposited dose. i) Ratio of backscattered to incident dose for the X junction.}
\label{img4}
\end{figure*}

The analysis of dose deposition by backscattered electrons provides crucial insights into the challenges associated with applying doses near the unexposed bridge region, which can lead to significant deformation due to the uneven spread of the applied dose. To mitigate the need for large doses in the vicinity of the junction area, it is essential to design geometries that strategically enhance the incidence of backscattered electrons. This approach aims to administer smaller, yet more uniform doses to the resist stack in the junction area from distributed regions. By comparing different geometries, we demonstrate how achieving a uniform dose distribution, or "saddle homogeneity," is key to enhancing the robustness of the process against variations.

To establish a reference for tolerance to variations in the fabrication process, we selected three different geometries to evaluate the dose-dependent room temperature resistance. The tested geometries are illustrated in Figure \ref{img3}a. We fabricated multiple samples with doses varying progressively from underdose to overdose, adjusting the design using proximity correction software. Standard test pads were created to minimize variations in process parameters, ensuring all junctions are produced within the same chip, subject to the same conditions. We applied a varying dose from 350 to 870 $\mu$C/cm$^2$, at 20 $\mu$C/cm$^2$ intervals. Each geometry started with a dose low enough to reveal the exposure pattern, but not high enough for the bridge structure to form, resulting in measured resistance equivalent to an open circuit. As we proceed to measure junctions exposed at higher doses, the bridge structure begins to form, and a measurable resistance emerges. At first resistance is very high because the bridge gap is small. This indicates the first junction measurements begin with the highest possible resistance and smallest junction area. As the dose increases it wears down both the PMMA and MMA, increasing the dose in unexposed area and widening the gap, leading to a decrease in measured resistance.  Once the dose deposited in the bridge region is high enough to compromise the PMMA bridge structure, it breaks and ceases to cast a shadow, allowing the top aluminum layer to form a closed circuit, culminating in a short measurement. In this manner we defined dose window starting from the first junction which resulted in a measurable resistance to the last. 

To analyze the results, we present in Table \ref{deltadose} the dose window within which measurable resistance was obtained for each analyzed geometry. For these geometries, the dose range for which a bridge structure forms is 260 $\pm$ 10 $\mu$C/cm$^2$ for the horseshoe and 160 $\pm$ 10 $\mu$C/cm$^2$ for the L junction. In contrast, the thin Dolan structure exhibits stability within a narrower span of only 20 $\pm$ 10 $\mu$C/cm$^2$ implying only one dose resulted a measurable resistance. Such a limited dose window for the thin Dolan design significantly reduces the process's success rate. Minor changes in temperature or development time could render the samples unusable. Furthermore we see that Horseshoe reproducibibility is at 96\% while the L junction 83\%. While  Although there is a notable difference in the standard deviation, the process was repeated in different facilities with similar reproducibility. We attribute this difference to the properties of the local EBL systems.

\begin{table*}[ht]
\centering
\caption{Experimentally defined reference for process tolerance per geometry. Reproducibility standards for geometries which have been thoroughly tested. Standard deviation for room temperature resistance. }
\begin{tabular}{|c|c|c|c|}
\hline
Junction Type & Supported dose variation($ \mu C / cm^2$)& Reproducibility &Std Deviation\\
\hline
Thin Dolan & 20 $\pm$ 10& Not reproducible&-   \\
\hline
L shape & 160 $\pm$ 10 & 83 $\%$ (20/24) & 3.2 $\%$   \\
\hline
horseshoe &  260 $\pm$ 10 & 96.3$\%$ (26/27) &  31.7$\%$  \\
\hline
\end{tabular}

\label{deltadose}
\end{table*}

To gain insight into the dose distribution across the junction area, we employed the Point Spread Function (PSF) \cite{Aya_1996}, by using the distributions derived from Monte Carlo simulations (Casino) and applying a MATLAB open source package Urpec \cite{urpec}. We then integrated PSF over the exposed geometries using a step size of 10 nm, we calculated the deposited dose within a 10x10 (100 $\mu$m$^2$) area, a close up is depicted in the dose distribution plot in Figure \ref{img3}b. A detailed examination of the unexposed bridge region is showcased in Figure \ref{img3}c, where orange lines represent vertical and horizontal traces without exposure. Further analysis of these traces enabled us to characterize the distribution of the total applied dose across various geometries and conditions. The noted asymmetry in the saddle-shaped energy distribution can be ascribed to one side of the junction being wider, facilitating a seamless contact between the upper and the oxidized (purple) lower layers of the Josephson junction, as demonstrated in the SEM angled image of the L-type junction (Figure \ref{img3}d), where the upper layer is observed making a clean and smooth contact with the lower layer.

To understand why the L and horseshoe are reproducible while the Thin Dolan design isn't, we begin by examining the ratio of energy deposited in the directly exposed areas to the indirectly exposed gap, shown in Figure \ref{img3}e. From this graph we can see that the exposed regions are subject to the same dose, however a sharp decrease in total deposited dose from 0.6 to less than 10$\%$ of this value for all geometries. A more detailed examination of the unexposed region, shown in Figure \ref{img3}f, indicates the total dose within 300 nm of the bridge section changes approximately 12 times for the most effective geometries, the horseshoe and L, and 20 times for the thin Dolan, which exhibits a lower success rate. Assuming the same dose is necessary for gap formation, the thin Dolan structure necessitates a higher geometry dose factor to modify the solubility the MMA layer, leading to pattern formation before the bridge structure, it is noticeable in the SEM image on Figure \ref{img3}g (left). This implies that a higher dose will also be deposited on the top PMMA layer, which is undesirable, as it weakens the bridge structure, causing some to break while others become narrower, thus reducing the reproducibility of Josephson junctions. To mitigate this issue, we investigate geometry-dependent designs to explore how the backscattered electron distribution to selectively modify the solubility the MMA resist.

Taking a closer look at the disposition of only the backscattered electrons for the 3 geometries (Figure 4a-b), we show that the Horse and L geometries have more homogeneous backscattered electron distribution than the Thin Dolan geometry. Furthermore, it can be noted that the Horse and L junction have over 50$\%$ more dose deposited over the bridge area. Showing that a smaller dose near the junction area is needed to form the bridge structure, preserving the PMMA structure for the reasons stated previously. This can be seen in Figure \ref{img3}g (right) where patterned resist for the horseshoe junction has a clear gap formed.

To further understand the impact of the incident electrons which may undergo forward scattering and drift from their point of incidence, we integrated the Point Spread Function (PSF) excluding the backscattered terms to analyze the distribution of incident energy and calculated the ratio of deposited dose between incident and backscattered electrons, denoted as $E_b/E_i$.
 Figure \ref{img4}c illustrates this ratio along the vertical orange line traversing the bridge area, as shown in Figure \ref{img3}c.
From the behaviour of $E_b/E_i$ in the bridge region for the three geometries observed, the contribution of backscattered electrons is at least twice as much as the dose deposited by incident electrons alone in the central part of the bridge region for the horseshoe and L-shaped designs. While, for the thin-Dolan design, this ratio in the vicinity of the bridge's central part exhibits a comparable effect.

Figure \ref{img4}d displays the total dose deposited by the incident beam in the bridge region for different base doses within the thin Dolan geometry. An increase in the base dose outside the bridge region elevates the total deposited dose within the bridge area as anticipated but does not alter the dose ratio between the regions outside the bridge and its central area. Consequently, a higher base dose intensifies the dose deposited throughout the entire resist stack, diminishing the bridge's stability.

\section{Discussion}

Simulations indicate that an optimized geometry should have enough exposed area within a 4 $\mu m$ radius of the junction area to eliminate the need of increased dose around the region of the bridge, which can causes the PMMA bridge structure to deteriorate. As a means to elucidate our findings pragmatically, we propose a innovative e-beam lithography technique in two steps. We suggest geometry should be optimized first rather than undercut. Once a a base dose for the geometry is found, strategic places are used to engineer the correct backscattered electron dose. A schematic of such a configuration is depicted on image Figure \ref{img4}e. The blue region receives the constant base dose, while the four green zones receive higher doses to create backscattered electrons. If sources were point-based and equidistant from the junction area, they would overlap with maximum interference over the unexposed junction region (Figure \ref{img4}f). Based on this logic we created four regions however to avoid extreme doses, we increased the area. Simulations showed that by raising the dose on the backscattered electron regions, the bottom of the vertical trace curve is affected with negligible effects on the border, efficiently raising the dose in the middle section of the bridge area (Figure \ref{img4}j). The ratio of backscattered to incident dose in Figure \ref{img4}i is significantly higher compared to previous junctions, and this geometry allows control over this ratio, providing an efficient means to precisely control the undercut to increase overall Josephson junction reproducibility. Although it has not been tested we believe designating zones to produce backscattered electrons is a promising means of increasing success rate in junction fabrication. 

\section {Conclusion}

This study provides a comprehensive investigation into the fabrication of Josephson junctions using 30 kV e-beam lithography.  Our  analyses highlight the critical role of backscattered electrons in the fabrication process, by carefully controlling their distribution. The study demonstrates that some methods achieve more uniform dose deposition than others, leading to improved junction stability and performance. To conclude our findings we propose a double dose unprecedented e-beam technique which separates, geometric exposure from undercut formation. This technique is a notable contribution to Josephson junction fabrication, but also many other lithography processes which require undercut engineering and control. As a means of implementation strategy, we provided a geometric example that takes advantage of simulated stability regions to apply a controlled homogeneous dose over the junction area. Introducing new features able to mitigate the challenges associated with the fabrication of bridge-like structures. These findings offer valuable insights into the fabrication of superconducting quantum devices and contribute to the advancement of quantum computing and sensing technologies. Josephson junctions are devices with great innovative potential, and our research democratizes their production, making it more accessible to fabricate Josephson junctions reliably using 30 kV e-beam systems, expanding access to groundbreaking technology for a broader community.

\vspace{6pt} 




\authorcontributions{All authors were involved in writing, reviewing and editing the manuscript.}

\dataavailability{Data available within the paper or via reasonable request. }

\funding{1) FAPERJ PROCESSO E-26/010.000980/2019 (NANO DEVICE NETWORK)\\
2) The electron microscopy/nanolithography work has been performed with the JSM 6490-LV / RAITH e-Line / JEM 2100F microscope(s) of the LABNANO/CBPF, Rio de Janeiro.\\
3) FAPERJ/CBPF/CNE project E26/200.830/2021\\
4) FAPERJ TEMATICO E-26/211.391/2021\\
5) LFDQ - Quantum Devices Physics Laboratory - FAPESP grant number 2017/08602-0 and 2021/01066-1, NAMITEC, National Council for Scientific and Technological Development – CNPq - 406193/2022-3\\
6)A. M. Rebello and N. Y. Klein, acknowledge FAPERJ and PETROBRAS for partially financing this project.\\
7)Scholarship for productivity and research process - CNPq- 305431/2022-6
}



\abbreviations{Abbreviations}{
The following abbreviations are used in this manuscript:\\

\noindent 
\begin{tabular}{@{}ll}
PEC & Proximity Effect Correction\\
UHV & Ultra High Vacuum\\
EBL & Electrom Beam Lithography\\
PMMA & Poly(methyl methacrylate)\\
MMA & Methyl methacrylate\\
    
\end{tabular}
}

\appendixtitles{yes} 
\appendixstart
\section{Appendix}
\subsection{Sample Preparation}
We will now describe how samples were fabricated and measured. Using optical lithography, metallic circuits with contact pads were created to test Josephson junctions at room temperature. For the e-beam bottom resist we chose a resist composed of co-polymers which have a 3 to 4 times higher sensitivity than PMMA resists. More specifically AR-P 617.08 (MMA), we apply a 500 nm coating by spinning at 4000 rpm. For the top e-beam resist we chose PMMA 950k to reach optimal thickness we chose 672.045 which comes out to 230 nm when spun at 4000 rpm. The first layer is baked at 200°C and the second at 180°C for 10 minutes each.

More specifically we engineered circuits consisting of 10 rows of 27 pads to be connected by a Josephson junction to a single trace for measurement. That enabled the probe testing of hundreds of Josephson junctions fabricated within a single sample to minimize variations. These were crafted using optical lithography on a Heidelberg DWL 66+. Our fabrication process involves depositing thin films, and selectively etching away materials by using wet etching for Aluminum or $SF_6$ RIE plasma etching for Niobium. Subsequently, employing a 30 kV EBL Dolan technique process on a Raith E-Line Plus to create resist structures for the fabrication of Josephson junctions. For room temperature measurements, we utilized a Lock-In with contact needle probing, while mK range measurements were conducted in a BlueFors DL400 Dilution Refrigerator.

\subsection{Experimental Application}
\begin{figure}[h!]
\resizebox{\hsize}{!}{\includegraphics[clip=true]{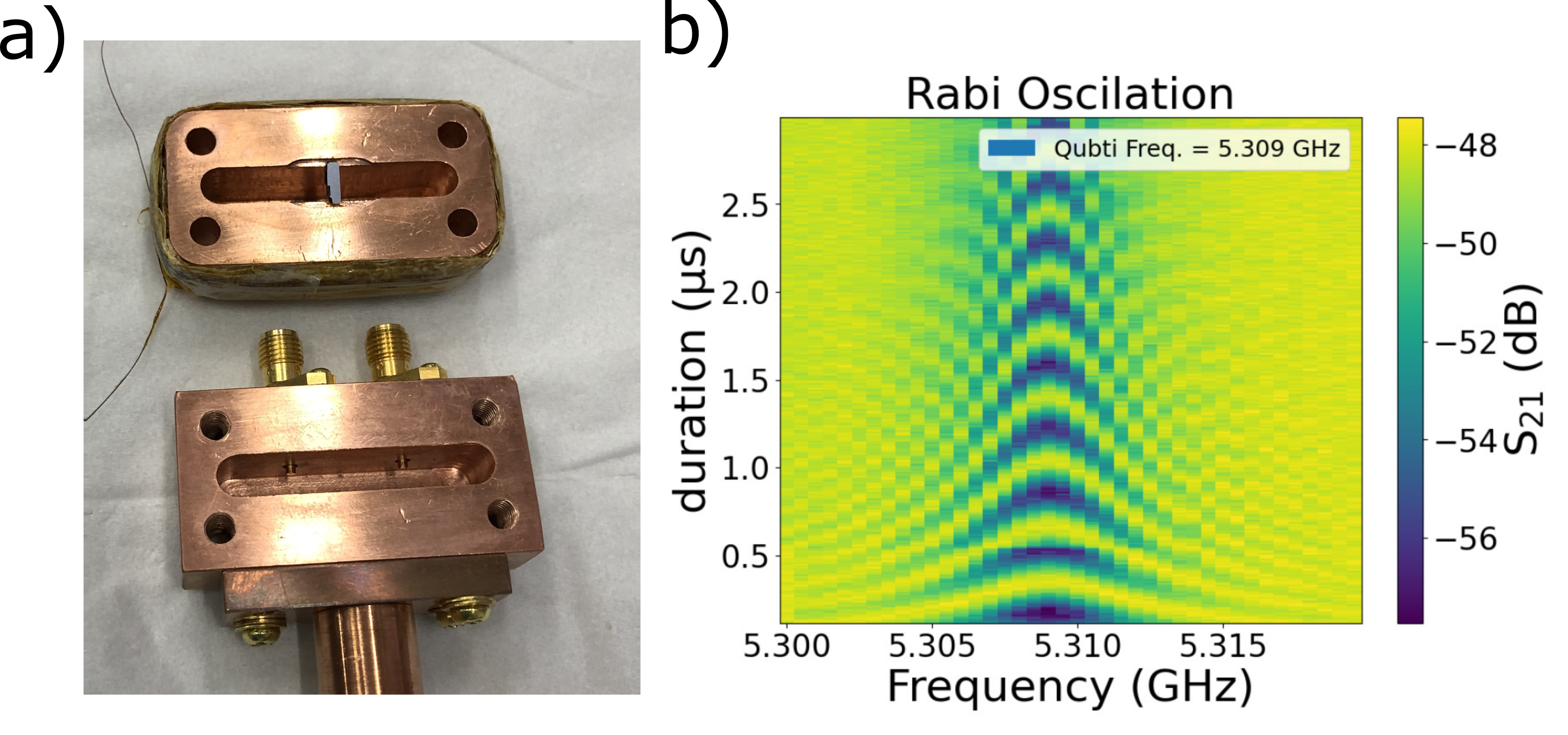}}
\caption{a) 3D copper cavity b) Quantum Rabi Map provides a profile of the dynamic evolution of qubit states in a time-dependent landscape.}
\label{img5}
\end{figure}
As a proof of concept, we successfully fabricated and analyzed a qubit within a (3D) cavity displayed on Figure \ref{img5}a. The qubit construction involved L-shaped junctions, similar to the ones shown in the scanning electron microscope (SEM) image in Figure \ref{img3}d. They were integrated into a squid with the rectangular en-looped area of (23x90) $\mu$m enclosed by two capacitive plates (300x800) $\mu$m. This device was placed in a 3D cavity and mounted onto the cold plate of a Dilution refrigerator.
The cavity and qubit frequency were measured at 7.389 GHz and 5.309 GHz respectively. In the course of our characterization, we observed Rabi Oscillations displayed on Figure \ref{img5}b and key performance metrics for the qubit. The relaxation time ($T_1$) was measured at 14.3 $\pm$ 0.4 $\mu s$ and the coherence time ($T_{2^*}$) was measured using the Ramsey protocol at 1.0 $\pm$ 0.1 $\mu s$. These characterizations provide evidence of the functionality and quality of the created Josephson junctions composing the qubit within the 3D cavity.
\begin{adjustwidth}{-\extralength}{0cm}

\reftitle{References}



\bibliography{aipsamp}

%


\end{adjustwidth}
\end{document}